\documentclass[twocolumn,showpacs,preprintnumbers,amsmath,amssymb]{revtex4}
\usepackage{graphicx}
\usepackage{dcolumn}
\usepackage{bm}

\begin{document}

%\preprint{APS/123-QED}

\title{Comment on ``Nonlinear band structure in Bose-Einstein condensates: Nonlinear Sch\"odinger equation with a Kronig Penney potential"}

\author{Ippei Danshita$^{1,2}$}
\email{danshita@nist.gov}
\author{Shunji Tsuchiya$^3$}
\address{$^1$National Institute of Standards and Technology, Gaithersburg, MD 20899, USA
\\
$^2$Department of Physics, Waseda University, 3-4-1 \=Okubo, Shinjuku, Tokyo 169-8555, Japan
\\
$^3$Dipartimento di Fisica, Universit\'a di Trento and CNR-INFM BEC Center, I-38050 Povo, Italy
}
\date{\today}

\begin{abstract}
In their recent paper [Phys. Rev. A {\bf 71}, 033622], B. T. Seaman {\it et al}.\! studied Bloch states of the condensate wave function in a Kronig-Penney potential and calculated the band structure.
They argued that the effective mass is always positive when a swallow-tail energy loop is present in the band structure.
In this comment, we reexamine their argument by actually calculating the effective mass.
It is found that there exists a region where the effective mass is negative even when a swallow-tail is present.
Based on this fact, we discuss the interpretation of swallow-tails in terms of superfluidity.

\end{abstract}

\pacs{03.75.Lm \ 05.30.Jn \ 03.75Kk}
\keywords{Bose-Einstein condensation, optical lattice, Kronog-Penney model, nonlinear Schr\"odinger equation}
\maketitle

%\protect \lowercase{} \textbackslash\textbackslash}
%%%%%%%%%%%%%%%%%%%%%%%%%%%%%%%%%%%%%%%%%%%%%%%%%%%%%%%%%%%%%%%%%%%%%%%%%%%%%%%%%%%%%%%%%%%%%%%%%%%%%%%%%%%%%%%%%%%%%%%%%%%%%%%%%%%%%%%%%%%%%%%%%%%%%%%%%%%%%%%%%%%%%%%%%%%%%%%%%%%%%%%%%%%%%%%%%%%%%%%%%%%%%%%%%%%%%%%%%%%%%%%%%%%%%%%%%%%%%%%%%%%%%%
%%%%%%%%%%%%%%%%%%%%%%%%%%%%%%%%%%%%%%%%%%%%%%%%%%%%%%%%%%%%%%%%%%%%%%%%%%%%%%%
%\begin{figure}[b]
%\includegraphics[scale=0.5]{tanh_pt}
%\caption{\label{fig:tanh}
%Schematic picture of the bosonic S-I-S junction.
%Two types of condensate wave functions are shown: the solid line shows the one %with CPPW and the dashed line shows the one with no CPPW.
%}
%\end{figure}
%%%%%%%%%%%%%%%%%%%%%%%%%%%%%%%%%%%%%%%%%%%%%%%%%%%%%%%%%%%%%%%%%%%%%%%%%%%%%%%
%%%%%%%%%%%%%%%%%%%%%%%%%%%%%%%%%%%%%%%%%%%%%%%%%%%%%%%%%%%%%%%%%%%%%%%%%%%%%%%%%%%%%%%%%%%%%%%           BdGE          %%%%%%%%%%%%%%%%%%%%%%%%%%%%%%%%%%%%%%%%%%%%%%%%%%%%%%%%%%%%%%%%%%%%%%%%%%%%%%%%%%%%%%%%%%%%%%%%%%%%%%%%%%%%%%%%%%%%%%%
%\section{MEAN FIELD THEORY AND MODEL POTENTIAL}
In their recent work on the band structure of Bose-Einstein condensates in a Kronig-Penney potential~\cite{rf:KP}, Seaman {\it et al}. have reported that when a swallow-tail energy loop is present in the band structure, the effective mass, which is known to be closely related to the dynamical instability of condensates in periodic potentials~\cite{rf:effmass}, is always positive and the condensate is dynamically stable in the lower potion of the swallow-tail.
In this comment, however, we recalculate the band structure by solving the time-independent Gross-Pitaevskii equation and find a region where the effective mass is negative in the lower portion of the swallow-tail.

We start from the time-independent Gross-Pitaevskii equation,
  \begin{eqnarray}
  \left[-\frac{1}{2m}\frac{d^2}{d x^2}-\mu+V(x)
  +g|\Psi_0(x)|^2\right]\Psi_0(x) =0,\label{eq:sGPE}
  \end{eqnarray}
with a Kronig-Penney potential,
  \begin{eqnarray}
  V(x)=V_0\sum_{j}\delta(x-ja),\label{eq:delta}
  \end{eqnarray}
where $a$ is the lattice constant, $V_0$ is the potential strength, $\mu$ is the chemical potential, $\Psi_0(x)$ is the condensate wave function, and $g$ is the coupling constant of the interatomic interaction.
Note that we set $\hbar=1$.
The normalization condition is given by
  \begin{eqnarray}
  \int_{ja}^{(j+1)a}\left|\Psi_0(x)\right|^2dx=N_{\rm C},\label{eq:normal}
  \end{eqnarray}
where $N_{\rm C}$ is the number of condensate atoms per site.
The energy of the condensate per site is given by
  \begin{eqnarray}
  E = \int_{ja}^{(j+1)a}dx\Psi_0^{\ast}
  \left[-\frac{1}{2m}\frac{d^2}{dx^2}+V(x)+\frac{g}{2}|\Psi_0|^2
  \right]\Psi_0.\label{eq:mean_ener}
  \end{eqnarray}

The condensate wave function can be written as $\Psi_0(x)=\sqrt{n_0}A(x){\rm e}^{{\rm i}S(x)}$, where $A(x)$ and $S(x)$ mean the amplitude and phase of the condensate. $A(x)$ is normalized by the density at the center of each site $n_0\equiv |\Psi_0\left(\left(j+1/2\right)a\right)|^2$.
Thus, Eq. \!(\ref{eq:sGPE}) is reduced to
   \begin{eqnarray}
      -\frac{1}{2m}\frac{d^2A}{dx^2}+\frac{Q^2}{2m}A^{-3}
      +V(x)A-\mu A+gn_0 A^3\!=\!0,      \label{eq:ampl}
   \end{eqnarray}
%\vspace{-8mm}
   \begin{eqnarray}
      A^2\frac{dS}{dx}=Q.\label{eq:conti}
   \end{eqnarray}
Equation \!(\ref{eq:conti}) is the equation of continuity and $Q$ describes the superfluid momentum.

%%%%%%%%%%%%%%%%%%%%%%%%%%%%%%%%%%%%%%%%%%%%%%%%%%%%%%%%%%%%%%%%%%%%%%%%%%%%%%%
\begin{figure}[b]
      \includegraphics[scale=0.48]{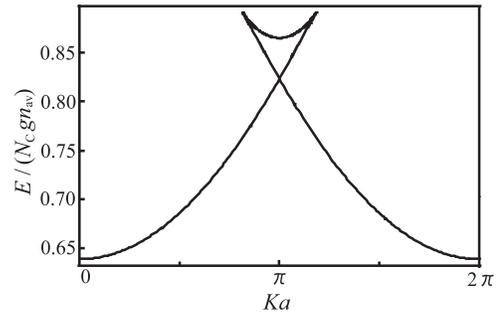}
\caption{\label{fig:1stband}
First Bloch band of the energy of the condensate $E(K)$ for $V_0=1 gn_{\rm av}\xi_{\rm av}$ and $a=5 \xi_{\rm av}$, where $n_{\rm av}=N_{\rm C}/a$ and $\xi_{\rm av}=(mgn_{\rm av})^{-1/2}$.}
   \end{figure}
%%%%%%%%%%%%%%%%%%%%%%%%%%%%%%%%%%%%%%%%%%%%%%%%%%%%%%%%%%%%%%%%%%%%%%%%%%%%%%%
%%%%%%%%%%%%%%%%%%%%%%%%%%%%%%%%%%%%%%%%%%%%%%%%%%%%%%%%%%%%%%%%%%%%%%%%%%%%%%%
\begin{figure}[thb]
      \includegraphics[scale=0.47]{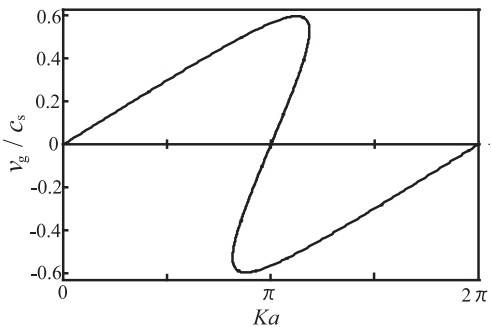}
\caption{\label{fig:group}
Group velocity $v_{\rm g}(K)$ for $V_0=1 gn_{\rm av}\xi_{\rm av}$ and $a=5 \xi_{\rm av}$, where $c_{\rm s}=(gn_{\rm av}/m)^{1/2}$.
}
   \end{figure}
%%%%%%%%%%%%%%%%%%%%%%%%%%%%%%%%%%%%%%%%%%%%%%%%%%%%%%%%%%%%%%%%%%%%%%%%%%%%%%%
%%%%%%%%%%%%%%%%%%%%%%%%%%%%%%%%%%%%%%%%%%%%%%%%%%%%%%%%%%%%%%%%%%%%%%%%%%%%%%%
\begin{figure}[t]
      \includegraphics[scale=0.48]{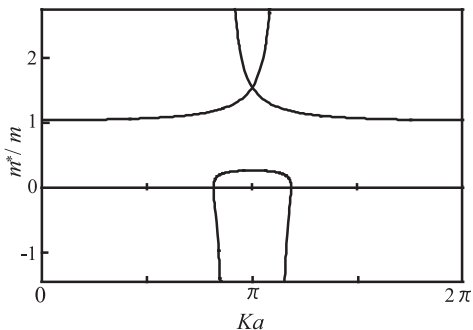}
\caption{\label{fig:mass}
Effective mass $m^{\ast}(K)$ for $V_0=1 gn_{\rm av}\xi_{\rm av}$ and $a=5 \xi_{\rm av}$.}
   \end{figure}
%%%%%%%%%%%%%%%%%%%%%%%%%%%%%%%%%%%%%%%%%%%%%%%%%%%%%%%%%%%%%%%%%%%%%%%%%%%%%%%
%%%%%%%%%%%%%%%%%%%%%%%%%%%%%%%%%%%%%%%%%%%%%%%%%%%%%%%%%%%%%%%%%%%%%%%%%%%%%%%
%\begin{figure}[t]
%      \includegraphics[scale=0.3]{PDsym}
%\caption{\label{fig:mass}
%Effective mass $m^{\ast}(K)$ for $V_0=1 gn_{\rm av}\xi_{\rm av}$ and $a=5 \xi_{\rm av}$.}
%   \end{figure}
%%%%%%%%%%%%%%%%%%%%%%%%%%%%%%%%%%%%%%%%%%%%%%%%%%%%%%%%%%%%%%%%%%%%%%%%%%%%%%%
Assuming that the condensate sits in the first Bloch band, one obtains the solution of Eq. \!(\ref{eq:ampl}) in the region $(j-1/2)a<x<(j+1/2)a$,
 \begin{eqnarray}
   \!\!\!\!\!\!&&A(x)^2=(1-\beta_{-})
   \nonumber\\
   \!\!\!\!\!\!&\times&\!\!\!
   {\rm sn}^2\!\left(\!\frac{\sqrt{\beta_{+}-\beta_{-}}
       (|x-ja|+x_0)}{\xi_0},
        \sqrt{\frac{1-\beta_{-}}{\beta_{+}-\beta_{-}}}\!\right)\!\!+\!
        \beta_{-},
        \label{eq:dens}
        %\nonumber
 \end{eqnarray}
where
  \begin{eqnarray}
  \beta_{\pm}\equiv 
  \frac{\mu}{gn_0}-\frac{1}{2}
  \pm\frac{1}{2}\sqrt{\left(\frac{2\mu}{gn_0}-1\right)^2-4(Q\xi_0)^2}.
  \end{eqnarray}
${\rm sn}(w,l)$ denotes the Jacobi elliptic sine function.
The healing length $\xi_0$ is expressed as $\xi_0=(m gn_0)^{-1/2}$.
$\mu$ and $x_0$ are determined as functions of $V_0$, $a$ and $Q$ by Eq. \!(\ref{eq:normal}) and the boundary condition at $x=ja$
  \begin{eqnarray}
  \left.\frac{dA^2}{dx}\right|_{ja+0}=
  \left.\frac{dA^2}{dx}\right|_{ja-0}+4mV_0 A^2(ja).
  \end{eqnarray}
The phase $S(x)$ can be calculated as 
  \begin{eqnarray}
  S(x)-S(ja)=\int_{ja}^{x}dx\frac{Q}{A^2}.
  \label{eq:phasedef}
  \end{eqnarray}
Detailed calculations of the condensate wave function Eqs. \!(\ref{eq:dens}) and (\ref{eq:phasedef}) have been shown in Ref.~\cite{rf:wareware}.

Imposing the Bloch theorem $\Psi_0(x+a)=\Psi_0(x){\rm e}^{{\rm i}Ka}$, the superfluid momentum $Q$ can be related to the quasi-momentum $K$ as
  \begin{eqnarray}
   Ka&=&S\left(\frac{a}{2}\right)-S\left(-\frac{a}{2}\right)
   \nonumber\\
     &=&\int_{-\frac{a}{2}}^{\frac{a}{2}}dx\, \frac{Q}{A^2},
   \label{eq:bloch}
  \end{eqnarray}
Substituting Eqs. \!(\ref{eq:dens}) and (\ref{eq:bloch}) into Eq. \!(\ref{eq:mean_ener}), one obtains the first Bloch band of the energy of the condensate $E(K)$ as shown in Fig. \!\ref{fig:1stband}.
A swallow-tail energy loop is present at the edge of the first Brillouin zone.

The group velocity $v_{\rm g}(K)$ and the effective mass $m^{\ast}(K)$ are given by
  \begin{eqnarray}
  v_{\rm g}(K)&=&\frac{\partial E}{\partial K},
  \\
  m^{\ast}(K)&=&\left(\frac{\partial^2 E}{\partial K^2}\right)^{-1}.
  \end{eqnarray}
$v_{\rm g}(K)$ and $m^{\ast}(K)$ are shown in Figs. \!\ref{fig:group} and \ref{fig:mass}, respectively.
Reflecting the presence of the swallow-tail, $v_{\rm g}(K)$ has a reentrant structure.
In the lower portion of the swallow-tail, $v_{\rm g}(K)$ increases as $K$ increases until $K_{\rm M}$, which gives the maximum value of the group velocity.
As $K$ increases further from $K_{\rm M}$ to the swallow-tail edge $K_{\rm E}$, $v_{\rm g}(K)$ decreases.
In the upper portion of the swallow-tail, $v_{\rm g}(K)$ increases as $K$ increases from $\frac{\pi}{a}$ to $K_{\rm E}$.
It is obvious in Fig. \!\ref{fig:mass} that the effective mass is negative in the region of the lower portion between $K_{\rm M}$ and $K_{\rm E}$.
This fact suggests the dynamical instability of the condensate in the region, which has been shown in Ref.~\cite{rf:wareware} by the linear stability analysis when the lattice constant is much larger than the healing length.
The region where the effective mass is negative is present also in the case of a sinusoidal periodic potential, because $v_{\rm g}(K)$ exhibits the same reentrant behavior~\cite{rf:diakonov}.
It is worth mentioning that the region of the negative effective mass exists for any values of $V_0$ and $a$ except in the limit of $V_0\rightarrow 0$.

In Refs.~\cite{rf:KP,rf:swallow}, the presence of swallow-tails in the band structure was interpreted as a manifestation of superfluidity of condensates.
This interpretation is inconsistent with the negative effective mass in the lower portion of the swallow-tail for the following reason.
It is assumed in the interpretation that the Bloch states in the lower portion correspond to local minima in the energy landscape.
However, they do not actually correspond to local minima, but to saddle points when the effective mass is negative.
Moreover, it was insisted in Refs.~\cite{rf:KP,rf:swallow} that as a superfluid, the condensate can screen out the periodic potential and the system follows the energy spectrum in the absence of the periodic potential until the superfluid momentum reaches the swallow-tail edge.
This argument is not valid, since the superfluidity breaks down due to the dynamical instability when the effective mass is negative.

I.D. is supported by a Grant-in-Aid from JSPS (Japan Society for the Promotion of Science).
%%%%%%%%%%%%%%%%%%%%%%%%%%%%%%%%%%%%%%%%%%%%%%%%%%%%%%%%%%%%%%%%%%%%%%%%%%%%
%\begin{acknowledgments}
%The authors would like to thank T. Kimura, S. Takei, S. Tsuchiya and T. Nikuni %for fruitful comments and discussions.
%They also greatly acknowledge useful comments of D. L. Kovrizhin.
%The work is partly supported by a Grant for The 21st Century COE Program (Holis%tic Research and Education center for Physics  of Self-organization 
%Systems) at Waseda University from the Ministry of Education, Sports, Culture, %Science and Technology of Japan.
%I. D. is supported by JSPS (Japan Society for the Promotion of Science) Researc%h Fellowship for Young Scientists.
%
%\end{acknowledgments}
%%%%%%%%%%%%%%%%%%%%%%%%%%%%%%%%%%%%%%%%%%%%%%%%%%%%%%%%%%%%%%%%%%%%%%%%%%%%%%%
%%%%%%%%%%%%%%%%%%%%%%%%%%%%%%%%%%%%%%%%%%%%%%%%%%%%%%%%%%%%%%%%%%%%%%%%%%%%%%%%%%%%%%%%%%%%%              REFERENCES                %%%%%%%%%%%%%%%%%%%%%%%%%%%%%%%%%%%%%%%%%%%%%%%%%%%%%%%%%%%%%%%%%%%%%%%%%%%%%%%%%%%%%%%%%%%%%%%%%%%%%%%%%%


\begin{thebibliography}{99}
\bibitem{rf:KP}
B. T. Seaman, L. D. Carr, and M. J. Holland, Phys. Rev. A {\bf 71}, 033622 (2005).
\bibitem{rf:effmass}
C. Menotti, A. Smerzi, and A. Trombettoni, New J. Phys. {\bf 5}, 112 (2003).

\bibitem{rf:wareware}
I. Danshita and S. Tsuchiya, Phys. Rev. A (to be published), e-print cond-mat/0610582.

\bibitem{rf:diakonov}
D. Diakonov, L. M. Jensen, C. J. Pethick, and H. Smith, Phys. Rev. A {\bf 66}, 013604 (2002).

\bibitem{rf:swallow}
E. J. Mueller, Phys. Rev. A {\bf 66}, 063603 (2002).

%\bibitem{rf:jacobi}
%H. T. Davis, {\it Introduction to Nonlinear Differential and Integral Equations} (Dover Publications, New York, 1962).

\end{thebibliography}
\end{document}